# Nonlocal phase-change metaoptics for reconfigurable nonvolatile image processing

*Guoce Yang[1], Mengyun Wang[1], June Sang Lee[1], Nikolaos Farmakidis[1], Joe Shields[2], Carlota Ruiz de Galarreta[2], Stuart Kendall[2], Jacopo Bertolotti[2], Andriy Moskalenko[1], Kairan Huang[1], Andrea Alù[3,4], C. David Wright[2]\*, Harish Bhaskaran[1]\**

[1]*Department of Materials, University of Oxford, Parks Road, OX1 3PH, United Kingdom*

[2]*Department of Engineering, University of Exeter, Exeter EX4 QF, United Kingdom*

[3]*Photonics Initiative, Advanced Science Research Center, City University of New York, New York USA*

[4]*Physics Program, Graduate Center, City University of New York, New York USA*

\*Email: david.wright@exeter.ac.uk and harish.bhaskaran@materials.ox.ac.uk

ABSTRACT

The next generation of smart imaging and vision systems will require compact and tunable optical computing hardware to perform high-speed and low-power image processing. These requirements are driving the development of computing metasurfaces to realize efficient front-end analog optical pre-processors, especially for edge-detection capability. Yet, there is still a lack of reconfigurable or programmable schemes, which may drastically enhance the impact of these devices at the system level. Here, we propose and experimentally demonstrate a reconfigurable flat optical image processor using low-loss phase-change nonlocal metasurfaces. The metasurface is configured to realize different transfer functions in spatial frequency space, when transitioning the phase-change material between its amorphous and crystalline phases. This enables edge detection and bright-field imaging modes on the same device. The metasurface is compatible with a large numerical aperture of ~0.5, making it suitable for high resolution coherent optical imaging microscopy. The concept of phase-change reconfigurable nonlocal metasurfaces may enable emerging applications of artificial intelligence-assisted imaging and vision devices with switchable multitasking.

**Introduction**

In the past decade, convolution neural networks (CNNs), a class of deep learning models for artificial intelligence (AI), have been widely used to solve target classification and recognition problems, with applications ranging from image-based medical diagnostics to real-time object detection for self-driving vehicles[1-4]. Practically, convolution layers - the fundamental front-end building blocks in CNNs – are responsible for most of the energy consumption, mainly due to the numerous multiplication operations they need to perform, which also ultimately slows down the processing speed in conventional electronic processors. Using light as carriers to process information has been a long-term vision due to the lower latency, smaller loss, and larger data throughput. Problems like relatively large footprints and scale-up issues make full optical solutions challenging but part substitution by photonic devices at the front end of a system is a feasible way to push the boundary of this field. There has been substantial progress in developing photonic integrated circuits[5,6], while alternatively free-space optics can manipulate input coherent light where two-dimensional information such as images is encoded directly, and process information without costly pre- and post-processing such as data vectorizations and electro-optical conversions. By operating directly in the spatial frequency domain, these devices can also dramatically speed up computation and reduce energy consumption compared to digital matrix mulitpliers.

A traditional optical method to process information in the spatial frequency space is to use a 4-f system[7], in which a lens transforms the spatial frequency space to the real space on the confocal plane and a spatially varying optical filter with position dependent transmittance is applied on that plane. This however makes the system bulky and prone to misalignments. In contrast, a "nonlocal" metasurface[8-15] - different from conventional "local" metasurfaces[16-22] that tailor the spatial wavefront by engineering the aperture response as a function of the spatial position for applications such as flat lenses and holograms - can apply Fourier operations on the input fields within an ultrathin form factor. This flat optics approach can be realized at the front end, with extremely small footprints, which makes it easier to integrate into compact optical systems. Due to these advantages, nonlocal metasurfaces are emerging as a promising platform to process information in the optical domain, with interesting applications such as image differentiators[23-27], space squeezers[28-30], light bullet generations[31] and reciprocal lenses[32] recently reported in the literature. However, such approaches so far have invariably used static metasurfaces that can perform a single function or processing task, limiting their impact and appeal for complex systems. Integrating multiple functionalities over a single nonlocal metasurface may bring significant processing advantages. In turn, this requires a nonlocal metasurface with capability for reconfiguration, which is the focus of the present work.

Very different from traditional local metasurfaces where the phase and amplitude response depend on meta-atoms locally and require individual control of meta-atoms, nonlocal metasurfaces have spatially uniform optical response, which makes globally switching metasurfaces enough to implement various functions. A reconfigurable nonlocal metasurface capable of image processing was recently proposed and demonstrated on a mechanical stretchable polydimethylsiloxane (PDMS) based platform[33,34]. Alternatively, phase-change materials (PCMs), the approach we use here, are good candidates to be applied for reconfigurable metasurfaces, due to their relatively large refractive index difference ($\Delta n > 0.7$) between their crystalline and amorphous states, and the non-volatility of such states, i.e., no energy consumption is required to remain in any particular state, although the switching operation itself consumes energy. Only recently have these materials been used to introduce reconfigurability in local metasurfaces, such as focus-varied metalens[35,36], beam steering devices[37,38], filtering[39] and more, raising significant interest in the area. The recent discovery of low-loss and lossless PCMs[40] and the substantial progress in developing integrated heaters capable of in-situ switching of PCM layers and meta-atoms[37,41-44], make them suitable for all-solid-state integrated meta-devices.

In this work, we theoretically propose and experimentally demonstrate a reconfigurable nonlocal metasurface made of $Sb_2Se_3$ - an optical lossless PCM in the near-infrared region. This phase-change nonlocal metasurface showcases a different transfer function in the spatial frequency domain when switched between its amorphous and crystalline phases, and, in the example demonstrated here, is able to perform edge detection and bright field imaging in the two distinct PCM phases. Our nonlocal metasurface is compatible with a numerical aperture (NA) of ~0.5, and a optical bandwidth of 50 nm corresponding to fractional bandwidth of 1/21 in theory, which enables applications in coherent optical imaging systems with high optical resolution. Furthermore, a dual-functionality coverslip based on the fabricated phase-change nonlocal metasurface (meta-coverslip) is demonstrated. The introduced concept of phase-change nonlocal metasurfaces may also enable all-solid-state dynamically tunable light modulators operating in the spatial frequency domain.

**Results**

A three-dimensional (3D) schematic of our conceptual phase-change nonlocal metasurface is shown in **Fig. 1a**. The metasurface consists of arrayed PCM nanopillars on a transparent substrate. With the nanopillars in different phase-states (amorphous or crystalline), the metasurface performs different functions. Here, the input image's edge is retrieved after light from the original image passes through the metasurface when the PCM is in the amorphous phase, i.e., edge detection mode, while no transformation happens when the PCM is in its crystalline phase, producing a bright field mode image. In this work, the nanopillars are realized by depositing optically low-loss PCM $Sb_2Se_3$ onto a sapphire substrate. The edge detection is implemented via a Laplacian operator in real space, which corresponds to a multiplication operator with a parabolic form in k-space, while the bright field mode corresponds to a pass-through-type transformation (shown in **Fig. 2b**). Therefore, in order to implement a dual function nonlocal metasurface, the transfer function $t(k_x, k_y)$ in k-space should be proportional to $k_x^2+k_y^2$ in the edge detection mode, while remaining flat ($t(k_x, k_y)=1$) in the bright field mode[13]. Our general method is to first design the metasurface working in one material phase in the edge detection mode which relies on a vertical oriented dipole resonance[13,24] to create a transmittance strongly dependent on incident angle, then to check whether transmittance is independent of incident angle in the other phase, so yielding the bright field mode. Thanks to the large refractive index change of phase change materials between the two phases, the resonance responsible for edge detection shifts far enough in wavelength when the material phase changes such that the dependence on incident angle disappears, making the metasurface work in the bright field mode.

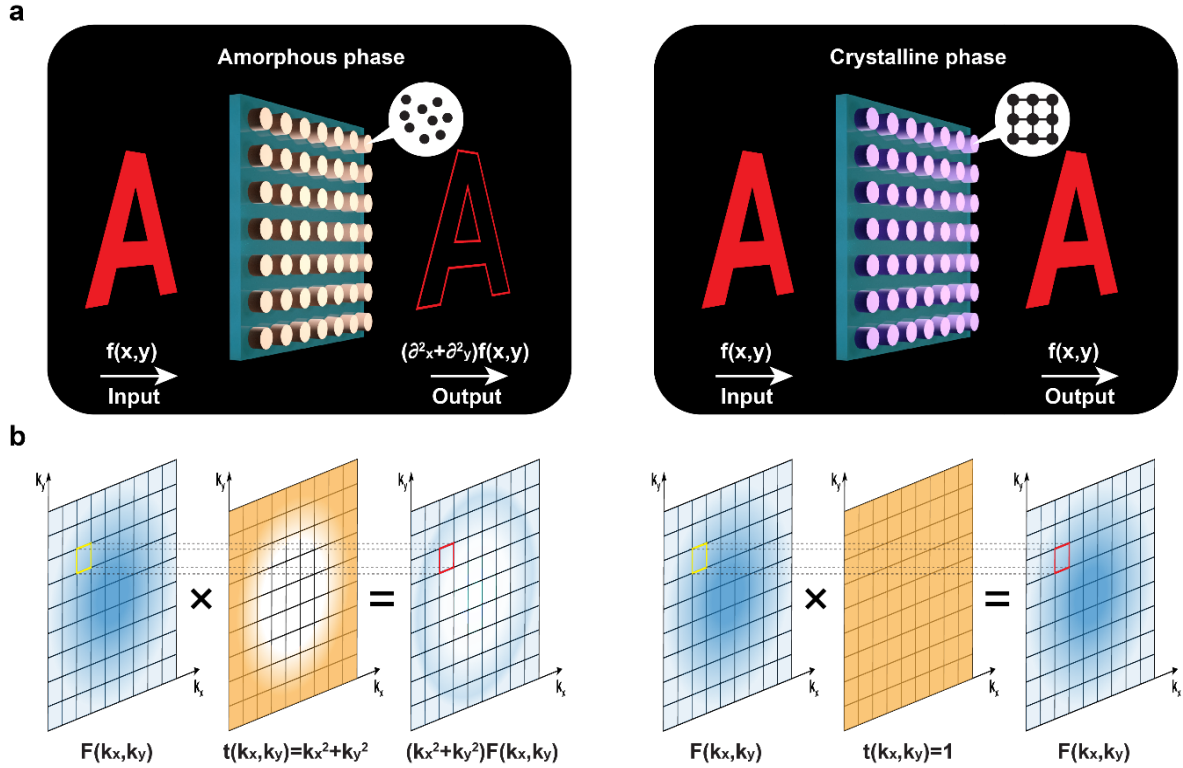

**Figure 1 Dual function phase-change nonlocal metasurface.** (a) 3D illustration of the phase change nonlocal metasurface performing the dual functions of edge detection in the amorphous phase and bright field imaging in the crystalline phase of $Sb_2Se_3$. (b) Schematic of the working principle of the metasurface in k-space, corresponding to the Fourier transform of the operation in real space.

Before optimizing the momentum-dependent (i.e. k-dependent) transmittance, we measured the complex refractive index of the amorphous and crystalline $Sb_2Se_3$ film using an ellipsometer. The measured results, shown in **Fig. 2a**, demonstrate that $Sb_2Se_3$ has low optical loss (imaginary part of the refractive index < 0.1) at wavelengths above 1000 nm in either phase, and the change of the real part of the refractive index between the two phases can be as large as 0.7 to 1 in the near-infrared region. In turn, the Mie resonant wavelengths of the nanopillars can shift significantly between the two phases, resulting in a large contrast of the transfer function, which give opportunities to implement different functions on a device with the fixed geoemtry. The 3D model of the nanopillar unit cell with the parameters of the geometric structure and the illumination configuration are shown in **Fig. 2b**.

We first design the metasurface working for edge detection in the amorphous phase and then check if the transfer function in the crystalline phase satisfies the bright field mode. The edge detection mode often requires incident angle dependent optical resonances, mostly featured by vertical oriented dipole modes or quasi bound states in the continuum [46-48]. We simulated the transmittance coefficient spectra of a periodic array of $Sb_2Se_3$ nanopillars - with sizes and periodicity specifically optimized to deliver edge-detection in the amorphous state and bright-field imaging in the crystalline state - as a function of incident angle with varied incident polarization states and structural phases of $Sb_2Se_3$. Here the transmittance coefficient is defined by the square root of the diffractive efficiency of the $0^{th}$ transmissive order. Given the subwavelength periodicity of our metasurface at the target wavelengths, no other propagating orders are excited. The diffraction efficiency of the metasurface was simulated by the rigorous coupled wave analysis (RCWA) method[45]. The simulated spectra exhibits a strong angle dependent contrast in the p-polarized incident state in the amorphous phase (**Fig. 2c**), which can be explained by the synergy of multiple resonances supported by the structure[34]. The in-plane electric and magnetic dipole resonances of the disks are close to each other in the spectrum at normal incidence, creating a band with almost zero transmittance from 980 nm to 1150 nm at normal incidence. With the increase of incidence angle, the in-plane resonance-induced spectral band is squeezed by another out-of-plane resonance mode, which causes more transmission for high spatial frequencies. By contrast, there is not much change in the spectra with increased incident angle in the crystalline phase, as shown in **Fig. 2d**. No significant change of transmittance as a function of incident angle was observed in the s-polarization state for either phase of the material. From the simulations, the metasurface behaves as a high-pass filter in spatial frequency from the wavelength of 1000 to 1150 nm in the amorphous phase and works as an all-pass filter from 1030 to 1080 nm in the crystalline phase. Therefore, our dual-function metasurface has 50 nm bandwidth in theory.

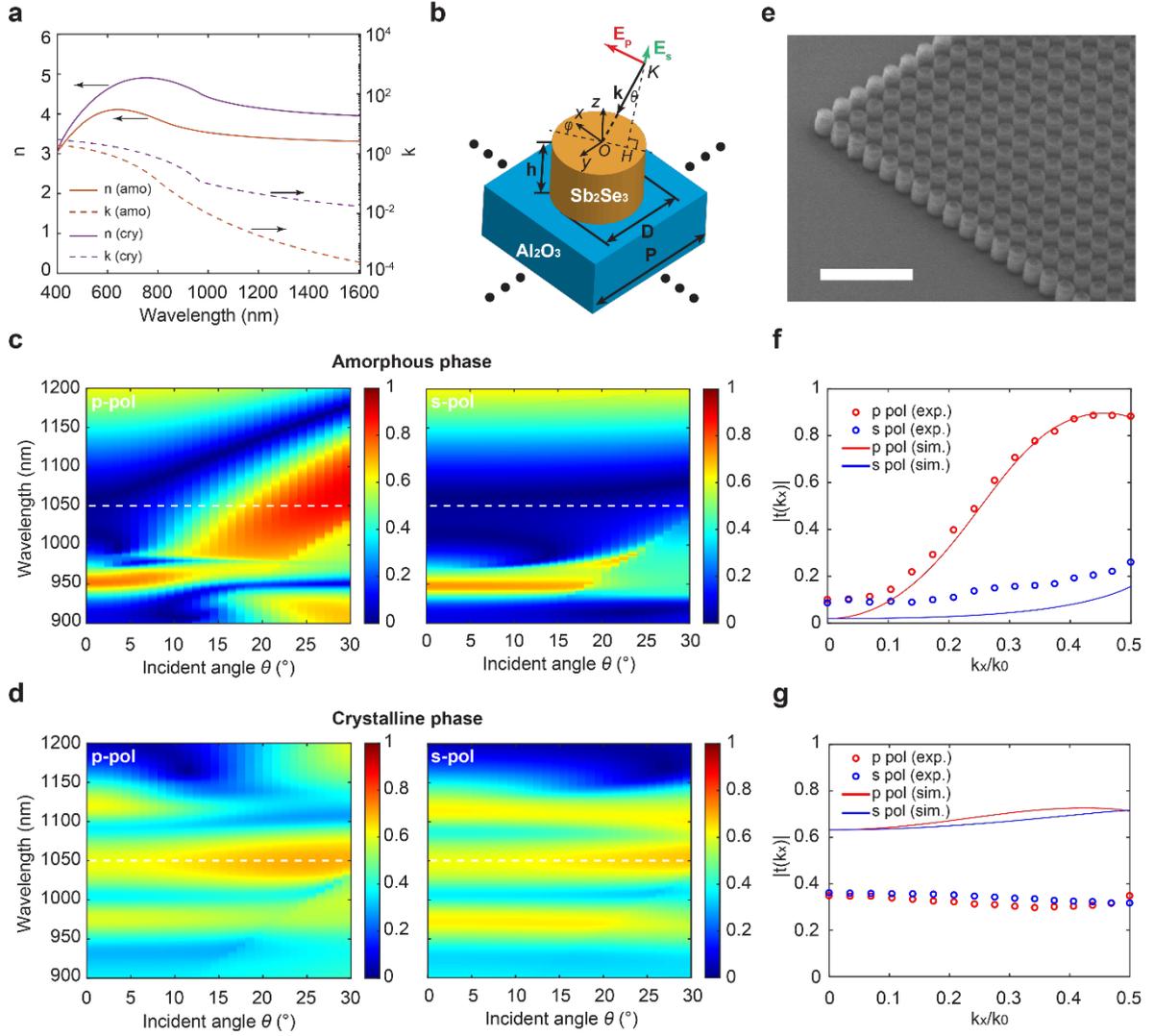

**Figure 2 Optical transmission properties of the nonlocal metasurface varying the $Sb_2Se_3$ phase.** (a) Measured refractive index of $Sb_2Se_3$ in the amorphous and crystalline phase. (b) 3D model of a unit cell of the metasurface with the configuration of incident light. $h$=300 nm, $D$=350 nm, and $P$=460 nm. (c, d) Simulated magnitude of transmittance coefficient spectra vs. incidence angle for the metasurface in the amorphous (c) and the crystalline phase (d), where the azimuthal angle $\varphi$=0°. (e) Representative SEM image of a fabricated metasurface. Scale bar: 2 μm. (f, g) Calculated and measured spatial frequency dependent transfer function of the metasurface in the amorphous (f) and crystalline phase (g) at the working wavelength of 1050 nm, corresponding to the dashed lines in (c) and (d).

We fabricated the metasurface using conventional processes, including $Sb_2Se_3$ film sputtering, electron beam lithography (EBL), and reactive ion etching (RIE) (**Methods**). The fabricated sample was finally characterized by scanning electron microscopy (SEM) and a representative captured image shows the high quality of the fabrication (**Fig. 2e**). To validate our design, we used a laser diode with emission wavelength of 1050 nm to measure the incident angle dependent the transfer function (**Fig. 2e** and **2f**). The transfer function is defined as the magnitude of transmittance coefficient, i.e. square root of the intensity transmittance which can be directly measured. The measurement setup is described in **Methods**. A strong spatial frequency dependent transfer function was experimentally demonstrated for the amorphous phase and p-polarized incidence, which coincides well with the simulation results. The amplitude of the transfer function reaches 0.9 when $k_x/k_0$ approaches 0.5, which means that the edge detection mode is efficiently (80%) compatible with an NA of 0.5. In the other cases (crystalline phase or s-polarization), the measured transfer function is not sensitive to the variation of the spatial frequency. There is an amplitude difference in the transfer function between measurements (0.65) and simulations (0.35) in the crystalline phase. This could be attributed to the scattering loss resulting from the increased non-uniformity of the crystallized metasurface (**Supplementary Information, Fig. S1**), and due to physical volume change which was already observed on crystallization in commonly used phase change materials and the element Se could make it more significant[49] (see also **Supplementary Information Fig. S2** and **Fig. S3**). In the crystalline phase and with p-polarization, the transfer function is flat as a function of spatial frequency, with an amplitude in the range 0.65 to 0.7 approximately, allowing us to image objects in bright field mode when the PCM nanopillars are in the crystalline state. More simulated transfer function of varied diameters are shown inf **Supplementary Information (Note C).**

We numerically analyze the 2D transfer function of our designed metasurface used for 2D image processing. We link the input electric field of the imaged object to the s- and p- polarized electric field components relative to the metasurface, and subsequently obtain the transfer function by introducing the simulated transmissive coefficients for s- and p- polarized incidence. The configuration of the input object and the metasurface is schematically shown in

**Fig. 3a**. Starting from the linear polarized light, we denote $E_{in}(x,y)\mathbf{u}$ as the input electric field of the imaged object in real space, where $\mathbf{u}$ is the unit vector along the polarization direction. The input electric field in k-space is denoted by $F_{in}(k_x,k_y)\mathbf{q}(k_x,k_y)$, where $F_{in}(k_x,k_y)$ is the Fourier transform of $E_{in}(x,y)$ and $\mathbf{q}(k_x,k_y)=(\mathbf{k}\times\mathbf{u})\times\mathbf{k}/|\mathbf{k}|$. Then we have the p- and s- polarized components: $E_p(k_x,k_y)=F_{in}(k_x,k_y)\mathbf{q}(k_x,k_y)\cdot\mathbf{e}_p$ and $E_s(k_x,k_y)=F_{in}(k_x,k_y)\mathbf{q}(k_x,k_y)\cdot\mathbf{e}_s$, where $\mathbf{e}_s=(\mathbf{k}\times\mathbf{z})/|\mathbf{k}|$ and $\mathbf{e}_p=(\mathbf{k}\times\mathbf{e}_s)/|\mathbf{k}|$ when $\mathbf{k}$ is not parallel with $\mathbf{z}$. For the special case of $\mathbf{k}//\mathbf{z}$, i.e. $k_x=k_y=0$, we set $\mathbf{e}_s=0$ and $\mathbf{e}_p=\mathbf{y}$ ($\mathbf{y}$ and $\mathbf{z}$ are the unit vectors along the $y$ and $z$ direction, respectively). Hence, the output electric field $\mathbf{F}_{out}(k_x,k_y)$ after the metasurface in the k-space can be generally written as[27]

$$\mathbf{F}_{out}(k_x, k_y) = F_{in}(k_x, k_y) \begin{pmatrix} \mathbf{e}_p \\ \mathbf{e}_s \end{pmatrix}^T \begin{pmatrix} t_{pp}(k_x,k_y) & t_{ps}(k_x,k_y) \\ t_{sp}(k_x,k_y) & t_{ss}(k_x,k_y) \end{pmatrix} \begin{pmatrix} \mathbf{q}\cdot\mathbf{e}_p \\ \mathbf{q}\cdot\mathbf{e}_s \end{pmatrix} \quad (1)$$

where the subscript p and s represents the p- and s-polarization, respectively. $t_{pp}$, $t_{ps}$, $t_{sp}$ and $t_{ss}$ are the transmission coefficients of the metasurface from one incident polarization to another output polarization. For example, $t_{ps}$ is the transmissive coefficient between the s-polarized incident light and the p-polarized transmitted light. The calculated transmission coefficients as the function of $k_x$ and $k_y$ in different phases are shown in **Fig. 3b**, **3c**, **3g** and **3h**. The cross-polarization coefficients $t_{ps}$ and $t_{sp}$ (the amplitude of ~0.2 corresponds to power transmittance of only 4%) are small enough compared with co-polarization coefficients $t_{pp}$ and $t_{ss}$ (the amplitude of ~0.9 corresponds to power transmittance of 81%) (**Supplementary Information, Note D**). Even though our sample was fabricated on sapphire substrate (c-cut), the optical axis is perpendicular to the ground plane, so the tiny birefringence effect of the substrate has no contribution to $t_{ps}$ and $t_{sp}$. Besides, the very slight anisotropy of the refractive index has negligible effect on the transmission profile in k-space (**Supplementary Information, Note E**). From the calculated results within the circle of NA<0.5 we approximately have

$$\begin{pmatrix} t_{pp}(k_x,k_y) & t_{sp}(k_x,k_y) \\ t_{ps}(k_x,k_y) & t_{ss}(k_x,k_y) \end{pmatrix} \propto \begin{cases} \begin{pmatrix} k_x^2+k_y^2 & 0 \\ 0 & 0 \end{pmatrix}, \text{amorphous} \\ \begin{pmatrix} 1 & 0 \\ 0 & 1 \end{pmatrix}, \text{crystalline} \end{cases} \quad (2)$$

This equation implies that our metasurface nearly blocks the s-polarized propagation and applies the Laplacian operator on p-polarized light. Besides, we define the transfer function of the metasurface as

$$t(k_x,k_y) = \left| \begin{pmatrix} \mathbf{e}_p \\ \mathbf{e}_s \end{pmatrix}^T \begin{pmatrix} t_{pp}(k_x,k_y) & t_{ps}(k_x,k_y) \\ t_{sp}(k_x,k_y) & t_{ss}(k_x,k_y) \end{pmatrix} \begin{pmatrix} \mathbf{q}\cdot\mathbf{e}_p \\ \mathbf{q}\cdot\mathbf{e}_s \end{pmatrix} \right| \quad (3)$$

which allows us to calculate the 2D transfer function of the imaged object illuminated by different polarizations of light via numerically simulated polarization dependent transmission coefficients. The calculated results for *x*-polarized, *y*-polarized and circular polarized incidence in the amorphous phase are shown in **Fig. 3d-3f**. The transfer function is close to $k_x^2$ but almost invariant with $k_y$ when the light is *x*-polarized, which means that the image is only differentiated along the *x* direction corresponding to the operator $\partial_x^2$. For the same reason, the *y*-polarized input light only induces differentiation along the *y* direction, i.e. $\partial_y^2$. In contrast, using circular polarized light makes the transfer function close to $k_x^2+k_y^2$, corresponding to differentiation along both directions. These optical properties of the metasurface enables detection of edges vertical to the polarization of the incident linear polarized light and full edge detection using circular polarized light[27]. We also calculated the transfer function of the metasurface in the crystalline phase at different polarizations in **Fig. 3i-3k**, which display near invariance to different k-vectors.

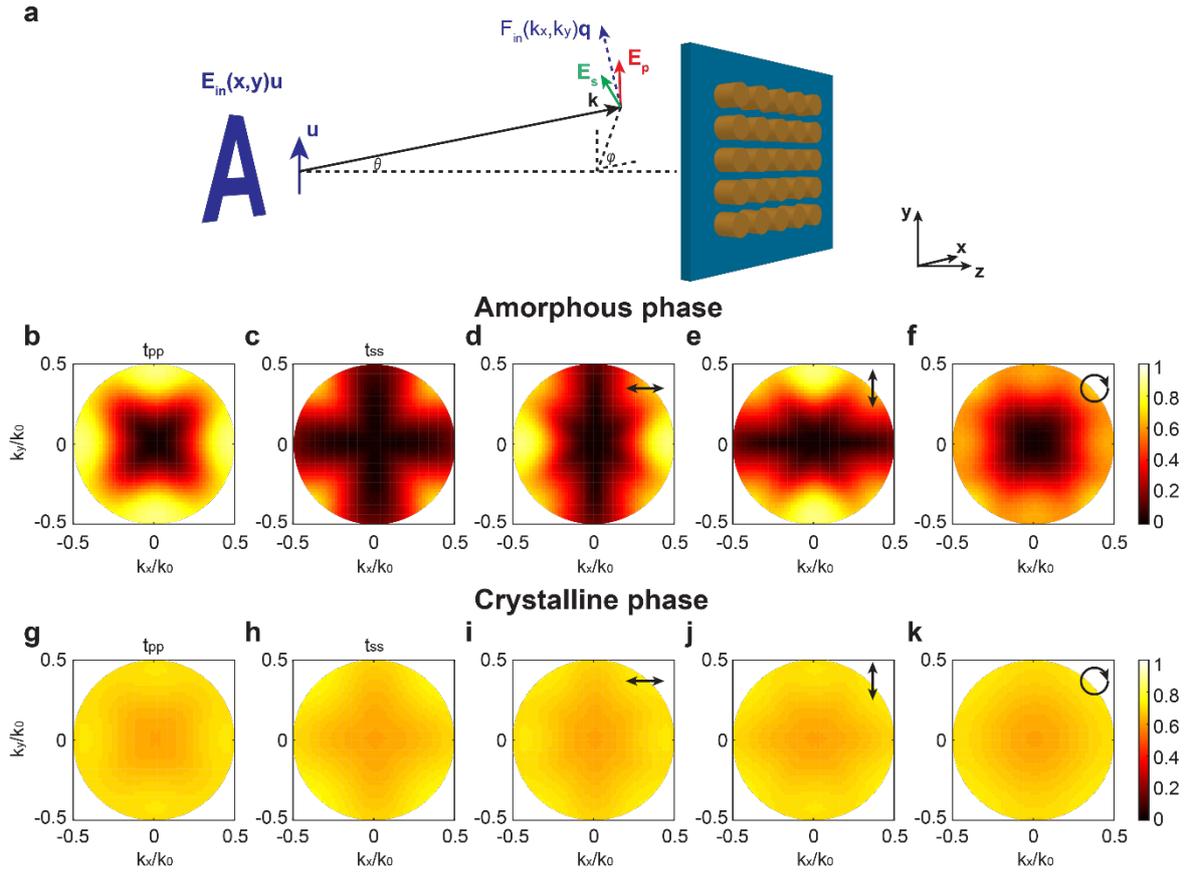

**Figure 3 Calculated 2D transfer function of the metasurface in the k-space.** (a) Schematic of relationship between the polarization state of the imaged object and the incident polarization state on the metasurface. (b,c) The calculated 2D transfer function for p-polarized (b) and s-polarized (c) incident light relative to the metasurface in the amorphouse phase. (d-f) The calculated 2D transfer function for different polarized illuminations on the samples: (d) x-polarized; (e) y-polarized; (f) circular polarized. The results of left and right hand circular polarized states are the same due to the symmetry of the metasurface. (g-k) The calculated 2D transfer function of the same configurations as in (b-f) but in the crystalline phase.

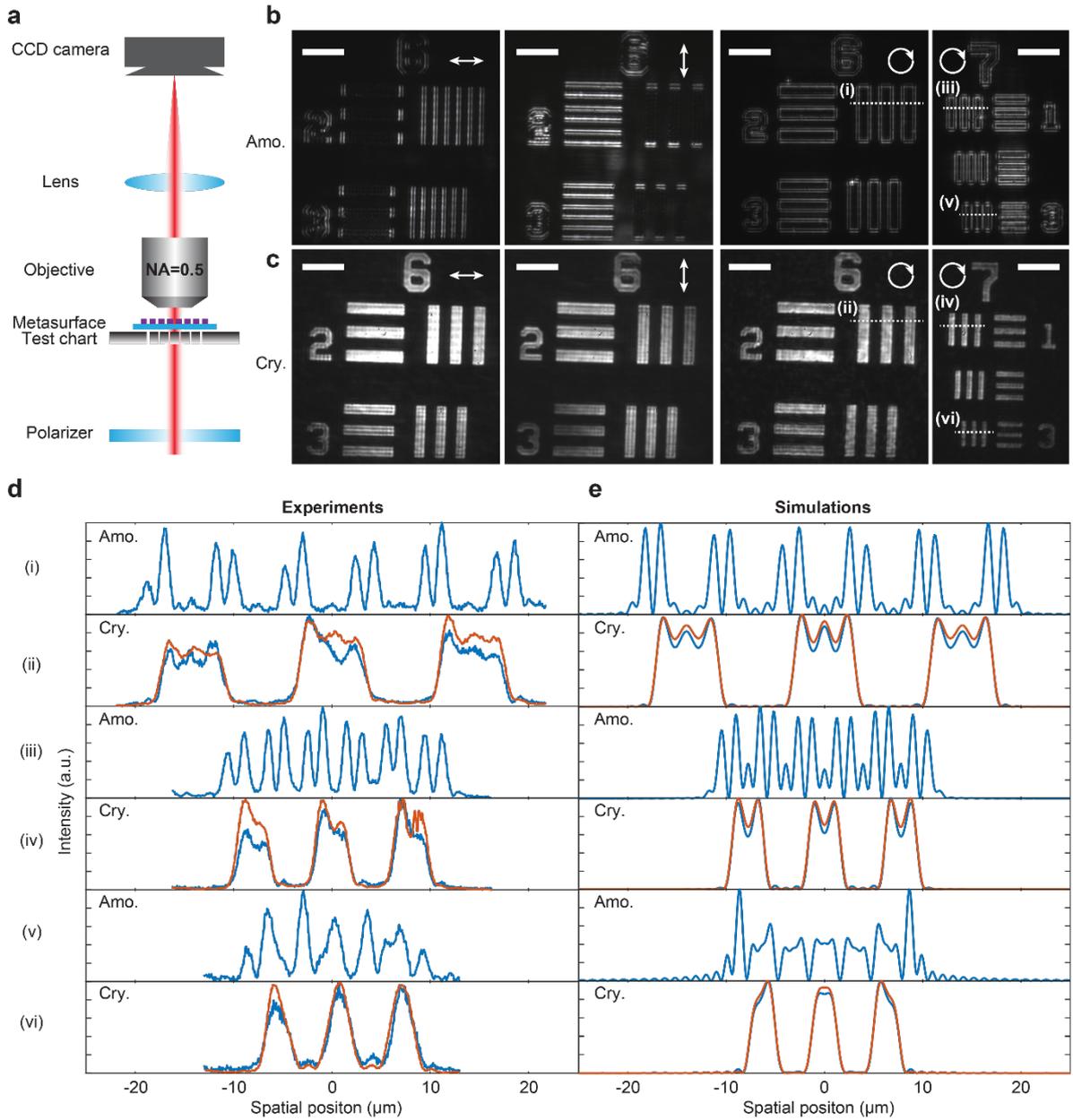

**Figure 4 Edge detection and bright field imaging modes.** (a) Schematic of the optical imaging setup in experiments, where the metasurface and the test chart are separated by ~0.1 mm. (b, c) Optical images of a negative 1951 USAF test target with the metasurface in the amorphous and crystalline phases. Different linear polarization orientations used in experiments are marked as arrows. Scale bar: 15 μm. (d, e) Measured and calculated light intensity profiles of the test target along the dashed lines marked in (b) and (c). Blue and red curves represent the results with and without the metasurface, respectively. All curves are normalized to the maximum of each.

Next, we experimentally demonstrate the phase-change reconfigurability of our nonlocal metasurface, supporting dual imaging functions, i.e., edge detection mode in the amorphous phase and bright field mode in the crystalline phase. The measurement setup is based on a purpose-built microscope schematically illustrated in **Fig. 4a**. A 1951 USAF negative resolution test target (used here as the imaged object) was illuminated by a diode laser beam with the wavelength of 1050 nm, and the metasurface was inserted between the test target and the objective to filter the spatial frequencies of the image before being collected by the imaging system. Test target imaging results show the good edge detection capability of the metasurface in the amorphous phase and the selectivity of enhanced edges depending on the incident polarization because of the polarization-sensitive transfer function shown in **Fig. 3**. The metasurface behaves like a left-right and an up-down edge detect kernel filter with horizontal and vertical polarization orientations, respectively, and full edge detection was implemented by illuminating a circular polarized light (**Fig. 4b**). To test the resolution of the edge detection operation, we imaged patterns with different group and element numbers and observed that the edge detection operation works well for feature sizes larger than 3.5 μm. Compared to the amorphous phase, the measured results from the crystalline $Sb_2Se_3$ metasurface shown in **Fig. 4c** show straightforward bright field imaging, with little differences, other than overall intensity reduction between the images shown and those without the metasurface (**Supplementary Information, Note F**).

In our experiments, we retrieved the grey values as the light intensity along the dashed lines on the captured images, for circular polarization, in **Fig. 4b** and **4c**, and plotted them in **Fig. 4d**. We also show in the figure the experimental results obtained with no metasurface present. First, we notice that there is no significant difference in the profile with and without the metasurface in the crystalline phase. The good overlap of the blue and red curves demonstrates the fidelity of the metasurface working in bright field mode, which works well for feature sizes as small as 2.2 μm, the samllest feature size on the test target used in experiments. More imaging results of smaller features are shown in the **Supplementary Information (Note G)**. For the bright field mode, the theoretical resolution limit of our system is 1.05 μm while it is out of the range of our resolution test chart. For the edge detection mode, the width of the obtained edge lines is also limited by the diffraction limitation of 1.05 μm, so it is difficult to clearly distinguish the edges when the feature size is smaller than around 4× diffraction limit (see **Supplementary Information, Note H**). When the metasurface operates in the edge detection mode, double lines appear at the edges of patterns due to the introduced Laplacian operator, which coincides well with theoretical results shown in **Fig. 4e**. In simulations, we first Fourier transform the real space profile to a spatial frequency space profile, then apply the simulated transfer function, and finally inversely Fourier transform it back to real space. Details are described in the **Supplementary Information (Note I)**. Besides, we also quantified the signal noise ratio of our edge detection results, which is >25 (**Supplementary Information, Note J**).

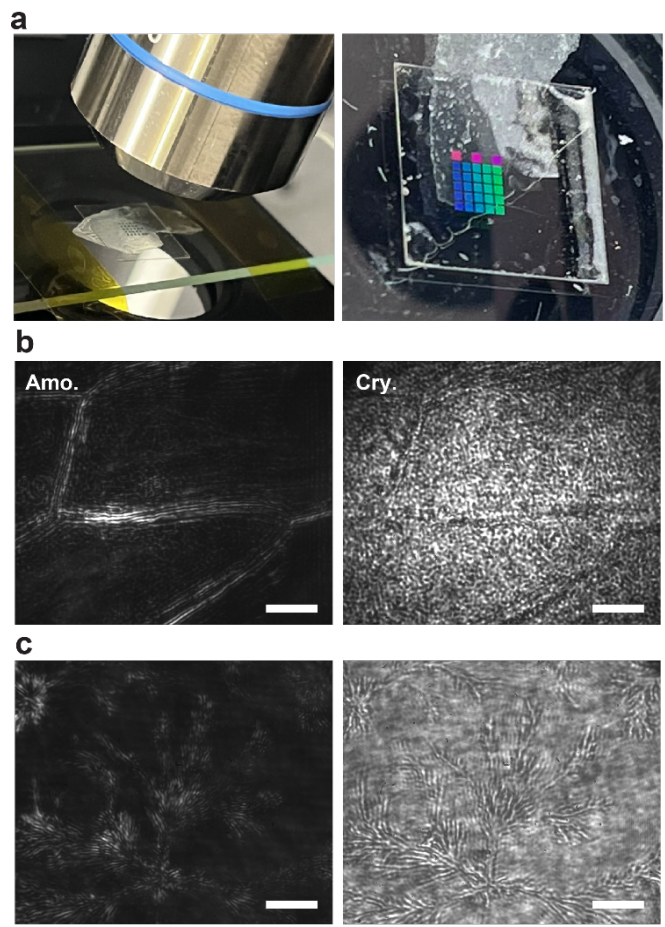

**Figure 5 A dual-functionality meta-coverslip for microscopy.** (a) A photograph of the phase-change nonlocal metasurface fabricated on a coverslip used to image onion epidermal cells. The different colored blocks are metasurfaces with different periodicities fabricated on the same substrate. (b) The imaging results for onion epidermal cells in the amorphous and crystalline phases of the metasurface. Scale bar: 15 μm. (c) The imaging results for thin sodium carbonate crystals self-grown on glass substrate. Scale bar: 15 μm.

Finally, we demonstrate a dual-functionality coverslip based on the phase-change nonlocal metasurface (meta-coverslip), as shown in **Fig. 5a**. The metasurface was fabricated on one side of an ultrathin coverslip (0.1 mm thick) and the other side of the coverslip was placed in contact with the object to be imaged (either onion epidermal cells or sodium carbonate crystals). The coverslip is made of sapphire, to ensure smaller deformation and less fragility during the fabrication process, but it can readily be replaced by other transparent substrates, such as cheap glass coverslips. The measurement setup and the illumination wavelength are the same as **Fig. 4**. **Figure 5b and 5c** shows the imaging results of onion epidermal cells and sodium carbonate crystals when the metasurface works in the edge detection and the bright field modes. The boundaries and shapes are quite clear when $Sb_2Se_3$ is amorphous, while they become much less identifiable when the metasurface is in the crystalline phase. Although in this work crystallization of $Sb_2Se_3$ is achieved by heating the metasurface on a hot plate at 225°C for 5 min and only one-way switching was shown, we stress that, ultimately,the there will be no necessity with the phase-change scheme to insert and remove the metasurface or apply an external force (as in the case of mechanically reconfigurable nonlocal metasurfaces) to switch between the bright field and edge detection microscopy modes ultimately. Instead, it is envisaged that in-situ switching of the metasurface using integrated microheaters, as demonstrated for example by Zhang et al.[37], could be implemented and make non-volatile cyclable switching possible.

**Discussion**

In conclusion, we have demonstrated a solid-state lossless phase-change nonlocal metasurface working as a reconfigurable spatial frequency filter. The metasurface can perform the dual function of bright field imaging and edge detection imaging. We have experimentally demonstrated one proof-of-concept application, a reconfigurable coverslip, for microscopy using our reconfigurable nonlocal metasurface. Our demonstration opens the door to active nonlocal metasurfaces targeting potential advanced applications such as dynamic spatial frequency filters for different image processing, tunable spacer squeezers[28-30] and active reciprocal lenses[32]. A parallel work using $VO_2$ as the PCM was recently reported[50]. $VO_2$ is a volatile PCM, requiring the temperature to be continuously held above threshold to keep the metasurface in one mode. By using a non-volatile PCM, we can eliminate power consumption while the function is being held once the PCM is already set or reset. The state of $VO_2$ only depends on the current temperature of the material, but the state of a non-volatile PCM such as $Sb_2Se_3$ is controlled by the past crystallization and amorphization process of the material, which makes dynamic switching non-trivial. Therefore, the next challenge is to address dynamic tunability with integrated heaters to make these metasurfaces switchable at the timescales of the crystallization and the amorphization of these active materials, something that has been achieved in reflective displays[51], but that may require significant further work for transmissive metasurfaces.

**Methods**

<u>Sample fabrication</u>

The 300 nm thick $Sb_2Se_3$ film was first sputtered on the clean sapphire substrate. Next a 300 nm thick negative tone electron beam resist (ma-N 2403) layer was spun coated on the $Sb_2Se_3$ film. The pattern was then transferred on the resist layer by EBL (JEOL JBX5500 50kV with an exposure dose of 250 μC/cm$^2$) and the developing process (developed in MF-319 for 1 min then rinsed by DI water). We used the fluorine-based RIE process (Oxford Instruments, $CHF_3$, 5 mT, 250 W) to etch the sample and finally removed the remaining resist by Microposit Remover 1165 to get the final device. The size of the fabricated metasurfaces is ~400×400 μm$^2$.

Optical measurements

The incident angle dependent intensity transmittance of the metasurface at the wavelength of 1050 nm was measured by the same purpose-built setup shown in Fig. 4a, but with a slight modification. The schematic setup is shown in **Supplementary Information (Note K)**. First, the laser beam from a fiber-coupled diode laser (Thorlabs, MCLS1) was collimated to free space with a beam diameter of ~3 mm, passed through a polarizer, and then slightly focused on the metasurface by a lens (f=200 mm) to make sure that the spot size is smaller than the area of the metasurface. Then, the transmissive light was collected by an objective with a small NA of 0.1 and coupled to a photodiode power sensor (Thorlabs, S122C). The use of a small NA objective is to make sure the working distance is large enough for sample rotation. The intensity transmittance is calculated by the ratio of the measured power with and without the metasurface.

**Acknowledgements**

This research was supported by European Union's Horizon 2020 research and innovation programme (Grant No. 101017237, PHOENICS Project) and European Union's EIC Pathfinder programme (Grant No. 101046878, HYBRAIN Project and No. 101098717 RESPITE Project). This research was funded in part by the UKRI [EP/T023899/1, EP/R001677/1 and EP/W022931/1]. A.A. acknowledges supports from the Air Force Office of Scientific Research MURI program.